\documentclass[pra,
twocolumn,
aps,
amsfonts,
amssymb,
nofootinbib,
byrevtex
,tightenlines,
preprintnumbers,
floatfix
]{revtex4}

\usepackage{amscd}
\usepackage{amsmath}
\usepackage{bm}
\usepackage[dvips]{color}                 %%% for colours in the graphs
\usepackage{graphicx}
\usepackage{eucal}
\usepackage{pslatex}

\def\){\right)} 
\def\({\left(} 
\def\]{\right]} 
\def\[{\left[}

\newcommand {\bea}{\begin{eqnarray}}
\newcommand {\eea}{\end{eqnarray}}
\newcommand {\be}{\begin{equation}}
\newcommand {\ee}{\end{equation}}

\def\Journal#1#2#3#4{{#1} {\bf #2}, #3 (#4)}

\def\PRL{\em Phys. Rev. Lett.}

\def\PRA{{\em Phys. Rev.} A}

\newcommand{\mcal}[1]{{\mathcal #1}}

\newcommand{\dpp}[1]{{\frac{d^3 #1}{(2\pi)^3}}}

\begin{document}
\title{Shear viscosity of a superfluid Fermi gas in the unitarity limit}

\author{Gautam Rupak
%\footnote{Email: {\tt grupak@u.washington.edu}} 
}
\email{grupak@u.washington.edu}
\author {Thomas Sch\"{a}fer}
\email{tmschaef@unity.ncsu.edu}
\affiliation{Department of Physics, North Carolina State University,
Raleigh, NC 27695}

\begin{abstract}
We compute the shear viscosity of a superfluid atomic Fermi 
gas in the unitarity limit. The unitarity limit is characterized
by a divergent scattering length between the atoms, and it 
has been argued that this will result in a very small viscosity. 
We show that in the low temperature $T$ limit the shear viscosity 
scales as $\xi^5/T^5$, where the universal parameter $\xi$
relates the chemical potential 
and the Fermi energy, $\mu=\xi \epsilon_F$. 
Combined with the high
temperature expansions of the viscosity our results suggest that
the viscosity has a minimum near the critical temperature $T_c$.
A na\"{i}ve extrapolation indicates that the minimum value of
the ratio of viscosity over entropy density is within a factor
of $\sim 5$ of the proposed bound $\eta/s\geq \hbar/{(4\pi k_B)}$.

\end{abstract}

\maketitle

%========================================================================
\section{introduction}
\label{intro}
%========================================================================
Shear viscosity $\eta$ can be defined as the shearing force
$F$ per unit area $A$ per unit velocity gradient in a laminar 
flow. For a flow in $x$-direction, with a velocity gradient 
$\nabla_y V_x$ in the $y$-direction 
\begin{align}\label{Newton_visc}
\frac{F}{A}=\eta\nabla_y V_x .
\end{align}
Viscosity relates the rate of momentum transfer  
to the velocity gradient. For dilute gases the microscopic mechanism 
for momentum transfer is provided by atomic collisions. This mechanism 
becomes more efficient as the mean free path gets larger because in
that case the atoms 
travel larger distances between collisions and transfer momenta between 
laminar layers of more disparate flow velocities. Thus viscosity $\eta$ 
is expected to be inversely proportional to the collision cross section 
$\sigma$. This leads to the question of whether there is a fundamental 
limit to how small the viscosity can get as the strength of the 
interaction is increased. Stated differently, we would like to 
determine the shear viscosity of the most ``perfect'' fluid.

 There is an old argument that suggests that quantum mechanics 
places a lower limit on the shear viscosity~\cite{Danielewicz:1984ww}. 
A rough estimate 
of the viscosity is provided by $\eta\sim n p \lambda$, where 
$n$ is the number density, $p$ is the average momentum, and 
$\lambda$ the mean free-path. Heisenberg's uncertainty principle 
requires $p\lambda \geq \hbar$ and the kinematic viscosity 
$\eta/n\gtrsim \hbar$. For relativistic systems particle number 
is not conserved and it is more natural to consider $\eta/s$,
where $s$ is the entropy density. As 
long as the entropy per particle is of the order $k_B$ we expect 
$\eta/s\gtrsim \hbar/k_B$.

 A new perspective on this idea is provided by a calculation, based 
on the AdS/CFT correspondence, of $\eta/s$ in the strong coupling 
limit of $N=4$ super-symmetric Yang Mills theory \cite{Policastro:2001yc}. 
This calculation gives $\eta/s=\hbar/(4\pi k_B)$, a value that is also 
obtained in other strongly coupled field theories that have a gravity 
dual. It is also known that the leading order correction to 
the limit of infinite coupling increases $\eta/s$. This has 
led to the conjecture that the strong coupling result is a 
universal lower bound for all fluids~\cite{Kovtun:2004de}:
\begin{align}
\label{ADS}
\frac{\eta}{s}\geq&\frac{\hbar}{4\pi k_B}.
\end{align} 
%This bound is satisfied by all known fluids.
 Liquid Helium 
comes to within an order of magnitude of the bound, and values
$\eta/s\sim (0.1-0.5)\hbar/k_B$ have been reported for the 
quark gluon plasma produced at RHIC \cite{Teaney:2003kp,Hirano:2005wx}.
There are suggestions in the literature that counter examples
can be found by considering non-relativistic systems for which 
the entropy per particle is very large ~\cite{Cohen:2007qr,Dobado:2007tm}, 
but currently no fluid that violates the bound is 
experimentally known.  

 An interesting system to study in this context is a cold atomic gas 
near a Feshbach resonance~\cite{JEThomas,TBourdel,KDieckmann,DJin,SJochim}.  
In $^6$Li and $^{40}$K gases, there exist hyperfine channels that 
support bound states. The magnetic moment of the bound state in 
these channels is different from the sum of the magnetic moments 
of the atoms that make the bound state. This allows one to use an 
external magnetic field to move the bound state energy relative 
to the continuum states, effectively making the bound state
arbitrarily shallow. In terms of scattering theory, a shallow bound
state corresponds to a large scattering length. At the Feshbach 
resonance, the atomic cross section is only limited by unitarity 
$\sigma(k)\sim 1/k^2$. The unitarity gas interaction is characterized 
by a divergent two-body scattering length $|a|\rightarrow\infty$ and 
a natural sized range $r\sim 1 \mathring{A}$. Even for a dilute gas 
with density $n\ll r^{-3}$, the unitarity gas with $|a|\rightarrow 
\infty$ is a strongly interacting system. In fact it is the
most strongly interacting non-relativistic system known, with a diverging
two-body collision cross section $\sigma(k=0)\sim a\rightarrow\pm\infty$.

The aim in this work is to improve the understanding of 
transport properties of the cold unitarity gas by performing 
a systematic calculation of the shear viscosity in the low
temperature superfluid phase. Combined with known results
in the high temperature limit \cite{Bruun} these results 
provide an estimate of the minimum viscosity. In the superfluid 
phase Cooper pairs break the $U(1)$ symmetry associated with 
the conservation of particle number. This implies that there 
is a Nambu-Goldstone boson, the phonon. At temperatures $T$ 
below the critical temperature $T_c$ for superfluidity, 
phonons dominate thermodynamic and transport properties of 
the system. 

 The paper is organized as follows. In Section~\ref{hydro} we
present the basic equations relating the shear viscosity to the 
phonon collision operator. The phonon interaction is derived in 
Section~\ref{phonon}, followed by a variational calculation of the 
viscosity in Section~\ref{Enskog}. A discussion of the result is 
presented. The discussion closely parallels the calculation of 
the viscosity in liquid Helium \cite{Khalatnikov,Maris} and, in 
particular, the CFL phase of dense quark matter \cite{Manuel:2004iv}. 
We end with the conclusions in Section~\ref{conclude}.

%========================================================================
\section{Transport Equation and Viscosity}
\label{hydro}
%========================================================================

 Viscosity as defined in Eq.~(\ref{Newton_visc}) is related to
internal stresses in a fluid. A more convenient definition is 
provided by the stress-energy tensor $T_{i j}$ of an almost 
ideal fluid. Close to equilibrium it can be expanded in 
derivatives of the flow velocity $V_i$,  
\begin{align}\label{del_T}
T_{i j}=&(P+\epsilon)V_i V_j -P\delta_{i j} +\delta
T_{i j}\ ,\\
\delta T_{i j}=&-\eta(\nabla_i V_j+\nabla_i
V_j-\frac{2}{3}\delta_{i j}\bm{\nabla}\cdot\bm{V})+\cdots ,\nonumber
\end{align}
where we only kept the traceless part of $\delta T_{ij}$. The trace
of $\delta T_{ij}$ is related to bulk viscosity. The ideal fluid
part of $T_{ij}$ is related to the thermodynamic variables pressure
$P$ and energy density $\epsilon$. In the superfluid phase the long
distance fluctuations of the order parameters and of the conserved 
quantities are described by the two-fluid hydrodynamics. The two 
components are a non-viscous superfluid, and a viscous normal 
fluid. The stress-energy tensor of the normal fluid is given by 
Eq.~(\ref{del_T}), where $V_i$ is now the velocity of the normal 
fluid. 

 If the normal fluid is composed of weakly interacting quasi-particles
the stress-energy 
tensor and the viscosity can be computed using kinetic theory. 
In the unitarity Fermi gas at very low temperature the quasi-particles 
are the phonons. The stress-energy tensor is given by ~\cite{Landau}
 \begin{align}
T_{i j}&=v^2\int\dpp{p} \frac{p_i p_j}{E_p} f_p,
\end{align}
where $f_p$ is the distribution function of the phonons with speed
$v$, momenta
$p_i$ and energy $E_p$. Close to the equilibrium $f_p=f_p^{(0)}+
\delta f_p$, where $f_p^{(0)}$ is the Bose-Einstein distribution 
and $\delta f_p$ is a small departure from equilibrium. Small
fluctuations can be parameterized in terms of departures of the 
thermodynamics variables $T,\mu,V_i$ from equilibrium, e.g.~$\delta f_p
\sim T\partial_{T}f_p^{(0)} \sim f_p^{(0)}(1+f_p^{(0)})/T$. This
motivates the definition $\delta f_p=-\chi(p) f_p^{(0)}(1+f_p^{(0)})
/T$ in terms of the unknown function $\chi(p)$. 
To project onto the shear stress, one uses the ansatz
\begin{align}\label{ansatz}
\chi(p) = &g(p) (p_i p_j-\frac{1}{3}\delta_{i j}p^2)
  (\nabla_i V_j+\nabla_j V_i
     -\frac{2}{3} \delta_{i j}\bm{\nabla}\cdot\bm{V}),
\end{align}
where only the traceless projection on the momenta $p$ is
relevant. Thus close to the equilibrium one can write
\begin{align}
\delta T_{i j}&=v^2\int\dpp{p} \frac{p_i p_j}{E_p} \delta f_p\\
&=- \frac{4v^2}{15 T }\int \frac{d^3
  p}{(2\pi)^3}\frac{p^4}{2 E_p}f_p^{(0)}(1+f_p^{(0)})
 g(p)\nonumber\\
&\times
(\nabla_i V_j+\nabla_j
V_i-\frac{2}{3} \delta_{i j}\bm{\nabla}\cdot \bm{V})\nonumber .
\end{align}
This determines the shear viscosity in terms of the function $g(p)$,
\begin{align}\label{viscosity}
\eta&=\frac{4v^2}{15 T }
\int \frac{d^3 p}{(2\pi)^3}
\frac{p^4}{2 E_p}f_p^{(0)}(1+f_p^{(0)}) g(p)\\
&= \frac{2v^2}{5 T }
\int \frac{d^3 p}{(2\pi)^3 2E_p}
f_p^{(0)}(1+f_p^{(0)})p_{i j} g(p)p_{i j}, \nonumber\\ 
p_{i j}&= p_i p_j-\frac{1}{3}\delta_{i j}
p^2
\nonumber .
\end{align}
The equation of motion for $g(p)$ is derived using the Boltzmann
equation 
\begin{align}\label{boltzmann}
\frac{d f_p}{dt}=&\frac{\partial f_p}{\partial t}+ \vec{v}\cdot
\vec{\nabla} f_p + \vec{F}\cdot \vec{\nabla_p}f_p= C[f_p],
\end{align}
relating the rate of change of the distribution function $f_p$ to the
collision operator  $C[f_p]$. In the absence of external force we take
$\vec{F}=0$. The left hand side of the relation Eq.~(\ref{boltzmann}) 
can be simplified further~\cite{Landau} to write:
\begin{align}
\label{LHSdf}
\frac{d f_p}{d f}\approx &
v\frac{f_p^{(0)}}{2 p T}(1+f_p^{(0)}) 
(p_i p_j -\frac{1}{3}\delta_{i j} p^2) \\
&\times(\nabla_i V_j+\nabla_j
V_i-\frac{2}{3} \delta_{i j}\bm{\nabla}\cdot \bm{V})
\ ,\nonumber\end{align}
where only the contribution relevant for shear viscosity was retained
in the Linear Response Approximation, leading order in the small deviation
from equilibrium.

\begin{figure}[t]
\includegraphics[width=0.45\textwidth,clip]{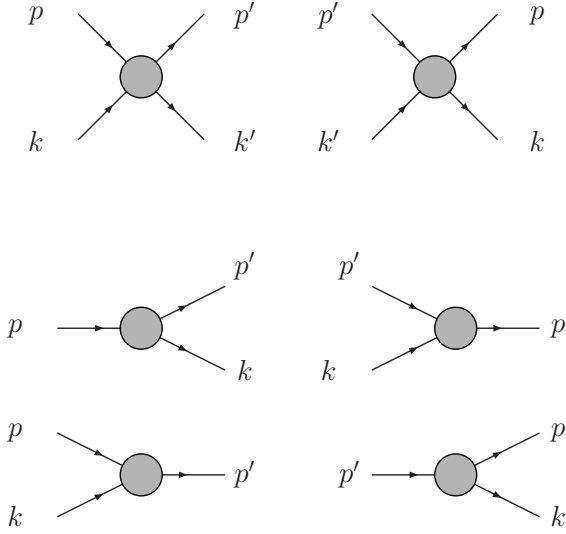}
\caption{\protect The first two diagrams contribute to the
  $2\leftrightarrow 2$ process, and the last four diagrams contribute
  to the $1\leftrightarrow 2$ processes. Only the leading order
  contribution to the shear viscosity $\eta$ from the vertices 
  are included.  
}
\label{collision}
\end{figure}

 Two types of contributions to the collision term $C[f_p]$ are typically 
considered: $(a)$ binary $2\leftrightarrow 2$ collisions in which the 
number of particles is conserved, and $(b)$ $1\rightarrow 2$ ``splitting'' 
processes in which the number of particles is not conserved. These 
processes are shown in Fig.~\ref{collision}. We will show in 
Section~\ref{phonon} that splitting processes do not contribute 
to shear viscosity at leading order in the low temperature approximation. 

%%%%%%%%%%%%%%%%%%%%%%%%%%%%%%%%%%%%%%%%%%%%%%%%%%%%%%%%%%%%%%%%%%%%
%%%%%%%%%%%%%%%%%%%%%%%%%%%%%%%%%%%%%%%%%%%%%%%%%%%%%%%%%%%%%%%%%%%%

 The  $2\leftrightarrow2$ collision integral is given by 
\begin{align}
C_{2\leftrightarrow2}[f_p]=&\frac{1}{2 E_p}\int \frac{d^3 k}{(2\pi)^3 2
  E_k}
\frac{d^3 k'}{(2\pi)^3 2 E_{k'}}
\frac{d^3 p'}{(2\pi)^3 2 E_{p'}}\\
&\times(2\pi)^4\delta^{(4)}(p+k-p'-k')|\mcal M|^2
D_{2\leftrightarrow2}, \nonumber 
\end{align}
where $D_{2\leftrightarrow 2}$ contains the distribution functions
and $|\mcal M|$ is the $2\leftrightarrow 2$ scattering amplitude 
shown in Fig.~\ref{collision}. The distribution functions are 
linearized in the small deviations from the equilibrium distribution,
$D_{2\leftrightarrow 2}\approx  D_{2\leftrightarrow 2}^{(0)}+
\delta D_{2\leftrightarrow 2}$. We find:
\begin{align}
\delta D_{2\leftrightarrow2}=& f_{k'}^{(0)} f_{p'}^{(0)} 
  (1+f_k^{(0)})(1+f_p^{(0)})\\
&\times\frac{\chi(p)+\chi(k)- \chi(p')-\chi(k')}{T}\nonumber, 
\end{align}
where we have used the equilibrium relation
\begin{align}
f_k^{(0)}f_p^{(0)}(1+f_{k'}^{(0)})(1+f_{p'}^{(0)})
=& (1+f_k^{(0)})(1+f_p^{(0)})f_{k'}^{(0)} f_{p'}^{(0)} \ .
\end{align}
This relation ensures that $D_{2\leftrightarrow2}^{(0)}=0$ and 
$C[f_p^{(0)}]=0$ in thermal equilibrium. 

 Using the ansatz in Eq.~(\ref{ansatz}) for $\chi(p)$, we get 
\begin{align}
\label{collisionA}
 C_{2\leftrightarrow2}[f_p]\approx& \frac{1+f_p^{(0)}}{2E_p T}
  \int \Gamma_{k; k' p'} (1+f_k^{(0)})f_{k'}^{(0)} f_{p'}^{(0)}\\
&\times
 \[g(p) p_{i j} +g(k) k_{i j}-g(k') {k'}_{i j} -g(p'){p'}_{ij}\]
   V_{i j}\nonumber \\
\equiv& F_{i j}[g(p)] V_{i j}\nonumber, 
\end{align}
where we have defined the linearized collision operator $F_{i j}[g(p)]$.
We have also defined
\begin{align}
\Gamma_{k; k' p'}=&\frac{d^3k}{(2\pi)^3 2 E_k}
\frac{d^3k'}{(2\pi)^3 2 E_{k'}}
\frac{d^3p'}{(2\pi)^3 2
  E_{p'}}\\
&\times(2\pi)^4\delta^{(4)}(p+k-k'-p')|\mcal M|^2 ,\nonumber\\
V_{i j}=&\partial_i V_i+\partial_j V_j-\frac{2}{3}\delta_{i j} V^2\
,\nonumber\\
 f_p^{(0)}=&\frac{1}{\exp(E_p/T)-1}\ .\nonumber
\end{align}
Using Eqs.~(\ref{boltzmann}), (\ref{LHSdf}) and (\ref{collisionA}), 
the linearized Boltzmann equation can be written as
\begin{align}
\label{linearBoltzmann}
 F_{ij}[g(p)]=&v\frac{f_p^{(0)}(1+f_p^{(0)})}{2 p T} p_{ij}. 
\end{align}
This result can be used to rewrite the relation for the viscosity 
in Eq.~(\ref{viscosity}) as 
\begin{align}
\label{viscosityFst}
\eta=&\frac{2}{5}\int\frac{d^3 p}{(2\pi)^3}p_{i j} g(p) F_{i
  j}[g(p)],
\end{align}
which will be useful later. We used the linear dispersion relation
$E_p=v p$ above, sufficient for the calculation as shown in the next
section. 
To complete the calculation of the 
solution to the collision equation we need to calculate the 
scattering amplitude $\sim|\mcal M|$ which we will turn to 
now.

%========================================================================
\section{Phonon Cross Section}
\label{phonon}
%========================================================================

 The phonon interaction for the unitarity gas in the 
superfluid phase can be derived from Galilean and gauge 
invariance~\cite{Greiter:1989qb,Son:2005rv}. Consider a 
microscopic Lagrangian for the  unitarity Fermi gas
\begin{align}
\mcal L_\psi = \psi^\dagger\[\partial_0 +\frac{\nabla^2}{2m}
+\mu\]\psi
-\frac{C_0}{4}(\psi^T\sigma_2\psi)^\dagger(\psi^T\sigma_2\psi),
\end{align} 
where $\psi$ is two component spinor, $m$ is the mass of the 
Fermion, $\sigma_2$ is the anti-symmetric Pauli matrix, and 
$C_0$ is an interaction strength that can be tuned to achieve
infinite scattering length. This Lagrangian is invariant under 
Galilean transformations, and under the gauge transformation $\psi\rightarrow
e^{i q(x)}\psi$ where the fictitious gauge field $A_\nu\rightarrow
A_\nu-\partial_\nu q$ is defined as $A_\nu=(\mu,\vec{0})$. We work 
in units where $\hbar=1=c=k_B$. 

 We require that the effective theory for the phonon field $\phi$ 
shares the symmetries of the microscopic Lagrangian. This implies
that the effective Lagrangian ${\cal L}_\phi$ is a function of 
\begin{align}
\chi =\mu-\partial_0\phi-(\bm{\nabla}\phi)^2/ (2m), 
\end{align} 
and its derivatives~\cite{Son:2002zn,Son:2005rv}. The functional
dependence on $\chi$ is further restricted by the observation
that the effective action at its minimum $\Gamma(\chi\!=\!\mu)= T \int d^3 x 
\mcal L_\phi$, for constant classical field $\partial_\nu \phi=0$, 
is equal to the pressure of the unitarity
gas. 
In the limit $|a|\rightarrow\infty$, $r=0$ which is nearly 
realized in cold atomic 
traps~\cite{JEThomas,TBourdel,KDieckmann,DJin,SJochim}, 
the unitarity gas is a scale 
invariant system. This implies that, up to a numerical
constant, the pressure $P$ has to be equal to that of the free
system. We write
\begin{align}
\label{P_uni}
 P = \frac{4\sqrt{2}m^{3/2}}{15\pi^2\xi^{3/2}}\mu^{5/2} ,
\end{align}
where the universal constant $\xi$ is sometimes called 
the Bertsch parameter in the nuclear physics community. 
Eq.~(\ref{P_uni}) implies $\mu=\xi \epsilon_F$, where 
$\epsilon_F=k_F^2/(2m)$, $k_F=(3\pi^2n)^{1/3}$, and $n$
is the number density. We conclude that~\cite{Son:2005rv} 
\begin{align}
\mcal L_\phi&= P(\mu\rightarrow \mu-\partial_0\phi
-\frac{(\bm{\nabla}\phi)^2}{2m}) +\mcal O(\partial_\mu\chi)\\
&= \frac{4\sqrt{2}\xi^{-3/2}
  m^{3/2}}{15\pi^2}\[\mu-\partial_0\phi
-\frac{(\bm{\nabla}\phi)^2}{2m}\]^{5/2}\nonumber+\ldots, 
\end{align}
where $\ldots$ corresponds to terms with derivatives of $\chi$. 
We can bring the kinetic term into the canonical form via a field 
rescaling $\phi\rightarrow \pi\xi^{3/4}\phi/[(m^3\mu)^{1/4}2^{1/4}]$.
We find expanding in derivatives of the phonon field, 
ignoring total derivatives of the dynamical field $\phi$ and constants
independent of $\phi$,  
\begin{align}
\label{Lphi}
\mcal L_\phi=& \frac{1}{2}(\partial_0\phi)^2 -\frac{1}{2}v^2
(\bm{\nabla}\phi)^2 \\
 & -\alpha\[(\partial_0\phi)^3
      -9v^2\partial_0\phi(\bm{\nabla}\phi)^2 \]\nonumber \\ 
 &-\frac{3}{2}\alpha^2\[(\partial_0\phi)^4
      +18v^2(\partial_0\phi)^2(\bm{\nabla} \phi)^2
      -27 v^4 (\bm{\nabla}\phi)^4\]+\cdots ,\nonumber
\end{align}
where $\alpha = \pi v^{3/2}\xi^{3/4}/(3^{1/4} 8\mu^2)$ and the 
Nambu-Goldstone boson speed is $v^2=2\mu/(3 m)$.

  The determination of $\xi$ is a non-perturbative many-body 
problem, and there are no exact analytical calculation available. 
Numerical calculations using fixed node Green's Function Monte Carlo
~\cite{Carlson2003,Chang2004,Astrakharchik2004}
or Euclidean lattice calculations
~\cite{Chen:2003vy,Lee:2005fk,Bulgac:2005pj,Burovski:2006} 
find $\xi\sim 0.3-0.4$. Our final result depends on this single
universal number $\xi$. 

 We can estimate the sizes of the different terms in the Lagrangian 
as follows: for the kinetic term to contribute to the generating 
functional its contribution should be $\mcal O(1)$ otherwise it 
will be damped in the exponential. Time derivatives scale as 
$\partial_0\sim T$, spatial derivatives as $\partial_i \sim T/v$,
and the volume integral scales as $d^4x\sim v^3/T^4$. This 
implies that $\phi\sim T/v^{3/2}$. We observe that the magnitude
of the phonon self coupling relative to the kinetic term scales
as $\alpha(\partial_0\phi)\sim \xi^{3/4}(T/\mu)^2$, a  
small correction for $T\ll \mu$.
Note that for a strongly interacting
unitarity gas  $T_c=(0.29\pm0.02)T_F\approx 0.7 \mu$
~\cite{TC} for $\xi=0.4$, which implies that $T_c$
is of the order $\mu$.

The Lagrangian in Eq.~(\ref{Lphi}) describes the leading order
phonon interaction for the processes shown in Fig.~\ref{collision}. At
this order, the phonon dispersion relation is linear with $E_p=
v|\bm{p}|$. Consequently, the splitting processes $1\leftrightarrow 2$
are collinear and cannot contribute to the shear viscosity.
  
%========================================================================
\subsection*{Binary Collisions}
%========================================================================

\begin{figure}[t]
\includegraphics[width=0.45\textwidth,clip]{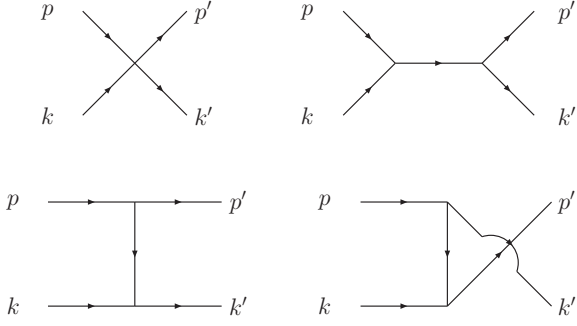}
\caption{\protect Leading order contributions to the binary collisions. 
}
\label{binary}
\end{figure}

 The leading order contribution to the binary collision processes 
in Fig.~\ref{collision} are shown in Fig.~\ref{binary}. The 
contribution of the four-phonon contact term to the scattering 
amplitude $\phi(p)+\phi(k)\to\phi(p')+\phi(k')$ is 
\begin{align}
i\mcal M_a&=-i\frac{3\sqrt{3}\pi^2 v^7\xi^{3/2}}{32\mu^4} k p k' 
  \{ 3\[\cos\gamma (k-6 k\cos\theta-3   p)\right. \nonumber\\
&\left.+k+p+(p-3 k) \cos\theta'+\cos\theta (k+p-6 p
   \cos\theta')\]\nonumber\\
&+3 \[3 \cos\theta-\cos\theta'+\cos\gamma 
(6 \cos\theta'-1)-1\] k'\nonumber\\
&+3 (\cos\gamma+\cos\theta+\cos
   \theta') p'+p'\},
\end{align}
where we have used $p+k=p'+k'$ and defined 
$\hat{\bm{p}}\cdot\hat{\bm{k}}  =\cos\theta$, 
$\hat{\bm{p}}\cdot\hat{\bm{k}'} =\cos\theta'$ and 
$\hat{\bm{k}}\cdot\hat{\bm{k}'} =\cos\gamma$. We also assumed
that the phonons are on-shell and that the dispersion relation 
is linear, $E_p=v |\bm{p}|$. Factors of $1/2$ from Bose symmetry
have been included in the amplitudes. 

%%%%%%%%%%%%%%%%%%%%%%%%%%%%%%%%%%%%%%%%%%%%%%%%%%%%%%%%%%%%%%%%%%%%
%%%%%%%%%%%%%%%%%%%%%%%%%%%%%%%%%%%%%%%%%%%%%%%%%%%%%%%%%%%%%%%%%%%%

 If the phonon dispersion relation is linear the $s$, $t$ and 
$u$-channel phonon exchange amplitudes diverge in the collinear 
limit. This corresponds to sub-sequent collinear splitting and
joining processes with an on-shell propagator in between. The 
collinear processes should not contribute to the shear viscosity,
but the numerical evaluation of collision integrals is more stable
if the infrared divergence due to the on-shell propagator is 
regularized by including the thermal damping of the phonon propagator. 
For this purpose we compute the imaginary part of the 
self-energy correction $\Sigma(p)$ to the phonon propagator.

%%%%%%%%%%%%%%%%%%%%%%%%%%%%%%%%%%%%%%%%%%%%%%%%%%%%%%%%%%%%%%%%%%%%
\begin{figure}[t]
\includegraphics[width=0.45\textwidth,clip]{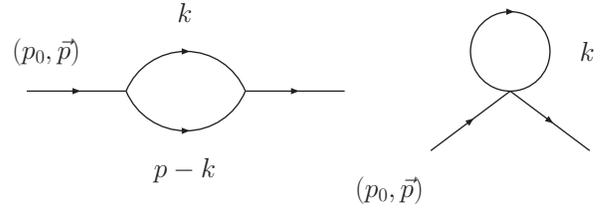}
\caption{\protect Leading order contributions to the phonon 
  self-energy correction. The tadpole does not contribute an 
  imaginary part.}
\label{selfenergy}
\end{figure}  
%%%%%%%%%%%%%%%%%%%%%%%%%%%%%%%%%%%%%%%%%%%%%%%%%%%%%%%%%%%%%%%%%%%%

 There are two self-energy diagrams at $\mcal O(\alpha^2)$,
 Fig.~\ref{selfenergy}. 
The 
tadpole graph does not generate an imaginary part and we only
compute the first diagram. We find
\begin{align}
\Sigma(p_0,\bm{p})
 &=\frac{\pi^2 v^3 \xi^{3/2}}{16\sqrt{3}\mu^4}T\sum_{n=-\infty}^\infty
 \int\frac{d^3 k}{(2\pi)^3}
 \frac{1}{\omega_n^2 +E_k^2}\\
&\times\frac{\[p_0 (2P\cdot K -K^2)+k_0 (P^2-2P\cdot K)\]^2}{(-i
  p_0+0^+-\omega_n)^2+E_{\bm{k-p}}}\nonumber .
\end{align}
The four-vector products  are defined as $P\cdot K= p_0 k_0-9v^2
\bm{p}\cdot \bm{k}$, $P^2=p_0^2-9v^2p^2$. This can be computed
following \cite{Manuel:2004iv}
and we find
\begin{align}
\label{sigma}
 \Sigma(p_0,\bm{p})&
   =-\frac{\pi^2 v^3\xi^{3/2}}{16\sqrt{3}\mu^4} \sum_{s_1,s_2=\pm}
    \int\frac{d^3 k}{(2\pi)^3}
    \frac{s_1 s_2}{4 E_k E_{\bm{p-k}}}\\
&\times\frac{1+f^{(0)}_{s_1 E_k}
  +f^{(0)}_{s_2 E_{\bm{p-k}}}}{ p_0+ i0^+ -s_1 E_k-s_2
    E_{\bm{p-k}}}\nonumber\\
&\times\[p_0 (2P\cdot K -K^2)+k_0 (P^2-2P\cdot K)\]^2\Big|_{k_0=s_1
  E_k}\nonumber. 
\end{align}  
The imaginary part of $\Sigma(p_0,\bm{p})$ arises from the pole
terms in the propagator. Analytic expressions for $\operatorname{Im}
\Sigma(p_0,\bm{p})$ can be found in Appendix~\ref{sigmaAnalytic}. For 
very time-like $|p_0|\gg |\bm{p}|$ external momenta 
\begin{align}
\operatorname{Im}\Sigma(p_0,\bm{p})&\approx
\frac{3\sqrt{3}\pi}{256}\xi^{3/2}
p_0^6
\[\frac{\exp(\frac{p_0}{2 T})+1}{\exp(\frac{p_0}{2 T})-1}\Theta(p_0)\right.\\
&\left.-
\frac{\exp(\frac{-p_0}{2 T})+1}{\exp(\frac{-p_0}{2 T})-1}\Theta(-p_0)
\].\nonumber
\end{align}
and for space-like $|p_0|\lesssim v |\bm{p}|$ external momenta with  
$v |\bm{p}|\ll T$
\begin{align}
\operatorname{Im}\, \Sigma(p_0, \bm{p})&\approx 
  \frac{2\sqrt{3}\pi^5}{5\mu^4 v}\xi^{3/2}T^4\frac{p_0^3}{p} 
  \Theta(v^2p^2-p_0^2). 
\end{align}
For the calculation these limiting forms provide sufficiently 
accurate representations of the exact one-loop expression 
in Eq.~(\ref{sigma}). We define the dressed phonon propagator
\begin{align}
i G(p_0, \bm{p})=\frac{i}{p_0^2-v^2\bm{p}^2
  +i \operatorname{Im}\Sigma(p_0,\bm{p})}. 
\end{align}
We can now collect the regularized $s$, $t$ and $u$-channel 
phonon exchange amplitudes. The $s$-channel amplitude is 
\begin{align}
i\mcal M_s&=- i\frac{\pi^2 v^5 \xi^{3/2}}{8\sqrt{3}\mu^4}(p+k)^2 G(p_0+k_0,
\bm{p+k})\\
&\times \[4 v^2 p k -P\cdot K\]\[4 v^2 p' k'-P'\cdot K'\]\nonumber .
\end{align} 
The $t$ and $u$-channel  amplitudes follow from crossing symmetry.
$i\mcal M_t=i\mcal M_s(k\leftrightarrow -p')$ and $i\mcal M_u = 
i\mcal M_s(k\leftrightarrow -k')$. We have 
\begin{align}
i\mcal M_t&=- i\frac{\pi^2 v^5 \xi^{3/2}}{8\sqrt{3}\mu^4}
   (p-p')^2 G(p_0-p'_0, \bm{p-p'})\\
 &\times \[4 v^2 p p' -P\cdot P'\]\[4 v^2 k k'-K\cdot K'\],\nonumber \\
i\mcal M_u&=- i\frac{\pi^2 v^5\xi^{3/2}}{8\sqrt{3}\mu^4}
   (p-k')^2 G(p_0-k_0', \bm{p-k'})\nonumber\\
 &\times \[4 v^2 p k' -P\cdot K'\]\[4 v^2 p' k-P'\cdot K\].\nonumber 
\end{align}

%========================================================================
\section{Variational Calculation}
\label{Enskog}
%========================================================================

  We are now in a position to compute the viscosity due to binary
collisions. We have to solve the linearized Boltzmann equation 
Eq.~(\ref{linearBoltzmann}) with the scattering amplitude determined
in the previous section, and then compute the viscosity using either
Eq.~(\ref{viscosity}) or Eq.~(\ref{viscosityFst}). This task is 
simplified by a number of useful properties of the linearized
collision operator $-F_{ij}[g(p)]$. The collision operator is a
linear operator on the space of functions $g(p)$. With a suitably
defined inner product this operator is hermitian and negative 
semi-definite. As a consequence it is possible to compute transport
properties using eigenfunction and variational methods~\cite{Resibois}.  

 We elect to use the trial functions 
\begin{align}
g(p)=p^n\sum_{s=0}^\infty b_s B_s(p),
\end{align}
where $n$ is a parameter that we choose for best 
convergence~\cite{Chen:2006ig}. 
The orthogonal polynomials $B_s(p)$ of order $s$ are defined such 
that the coefficient of the highest power $p^s$ is 1 and that 
the orthogonality conditions~\cite{Dobado:2001jf}  
\begin{align}
\int\frac{d^3 p}{(2\pi)^3 } p_{i j} 
  \frac{f_p^{(0)} }{2E_p}(1+f_p^{(0)}) p_{i j} p^n B_r(p) B_s(p)=& 
  A_{r s}\delta_{r s}, 
\end{align} 
are satisfied. Starting from $B_0=1$ we can recursively determine
all the $B_s(p)$. This also defines the normalization factors $A_{r
  s}$. The polynomials $B_s(p)$ are a generalization of 
the Sonine (modified Legendre) polynomials to Bose-Einstein statistics
and linear dispersion relations. 

 Inserting the trial function into Eq.~(\ref{viscosity}) we find the
following expression for the viscosity 
\begin{align}
\label{eta_b0}
\eta[g(p)] =&  \frac{2v^2}{5 T }\sum_{s=0}^\infty b_s 
\int \frac{d^3 p}{(2\pi)^3 2 E_p}
f_p^{(0)}(1+f_p^{(0)})\\
&\times p_{i j} p^n p_{i j} B_s(p)\nonumber\\
=& \frac{2v^2}{5 T }\sum_{s=0}^\infty b_s A_{0s}\delta_{0s}
=\frac{2v^2}{5 T } b_0
A_{00}\nonumber. 
\end{align}
Alternatively, we can use the trial function in Eq.~(\ref{viscosityFst}).
We get
\begin{align}
 \label{phase4}
\eta[g(p)] = &\frac{2}{5}\int\frac{d^3 p}{(2\pi)^3}p_{i j} g(p) F_{ij}[g(p)]
   \equiv \sum_{s, t=0}^\infty b_s b_t M_{s t},
\end{align}
where $M_{st}$ are the matrix elements of the linearized collision 
operator
\begin{align} 
M_{st}=& \frac{2}{5 T}\int d\Gamma_{pk;p'k'}
(1+f_p^{(0)})(1+f_k^{(0)}) f_{p'}^{(0)} f_{k'}^{(0)}\\
&\times p^n B_s(p) 
p_{i  j}
 \[B_t(p) p^n  p_{i j} + B_t(k) k^n k_{ij} \right.\nonumber\\
&\hspace{0.2in}\left. -B_t(p') {p'}^n p'_{i j} 
-B_t(k') {k'}^n k'_{i j}\]\nonumber ,
\end{align} 
with the four-particle phase space factor
\begin{align}
\Gamma_{pk;p'k'}
=&\frac{d^3p}{(2\pi)^3 2 E_p}
\frac{d^3k}{(2\pi)^3 2 E_k}
\frac{d^3k'}{(2\pi)^3 2 E_{k'}}
\frac{d^3p'}{(2\pi)^3 2
  E_{p'}}\\
&\times(2\pi)^4\delta^{(4)}(p+k-k'-p')|\mcal M|^2.\nonumber
\end{align}
Eqs.~(\ref{eta_b0}) and (\ref{phase4}) are consistent provided
\begin{align}
\label{consis}
\sum_{t=0}^\infty M_{s t} b_t =&\frac{2 v^2}{5 T} A_{00}\delta_{s 0} .
\end{align}
This is a simple linear equation for $b_i$ which is solved by
\begin{align}
\(\begin{array}{c}b_0\\b_1\\b_2\\ \vdots \end{array}\)
= &\frac{2 v^2}{5 T} A_{00} M^{-1}\cdot
\(\begin{array}{c}1\\0\\0\\ \vdots\end{array}\).
\end{align}
Once $b_0$ is determined we can extract the viscosity from 
Eq.~(\ref{eta_b0}). In practice we pick a value for $n$ and
study convergence as the number of orthogonal polynomials is
increased. What is nice about the method is that this is a 
variational procedure. One can show that~\cite{Resibois} 
\begin{align}
\eta \geq \frac{4v^4}{25 T^2} 
  \frac{(b_0A_{00})^2}{\sum_{s, t} b_s b_t M_{s t}}
\end{align}
for any $n$ and sets of $b_s$. The condition that the bound is 
optimized with respect to the expansion coefficients $b_s$ is 
equivalent to the consistency condition Eq.~(\ref{consis}). 

 The scaling behavior of the viscosity with respect to the
temperature and the Bertsch parameter $\xi$ is easily derived. 
We scale all momenta as $p\rightarrow Tp/v$. Using $E_p=v |\bm{p}|$ 
this fixes the scaling of the energies. All terms in the scattering 
amplitude $\mcal M$ have the same scaling behavior, except for
a sub-leading correction due to the self-energy insertion in 
the phonon propagator. In terms of scaled momenta the phonon
propagator $G(p_0,p)$ can be written as 
\begin{align}
iG(p_0,\bm{p}) =\frac{1}{T^2}\frac{i}{p_0^2 -p^2+i \xi^{3/2}(T/\mu)^4
\operatorname{Im}\hat{\Sigma}(p_0,\bm{p})}. 
\end{align}
The scaling of the scattering cross section is $|\mcal M|^2\sim \xi^{3}
v^6(T/\mu)^8$, and the self energy term induces corrections that 
are functions of $\xi^{3/2}(T/\mu)^4$. We find
\begin{align}
A_{00}=&\frac{T^{6+n}}{v^{7+n}}\hat{A}_{00},\\
M_{st}=&\frac{T^{2n+15+s+t}}{v^{2n+7+s+t}}\frac{\xi^3}{\mu^8}
 \hat{M}_{st},\nonumber
\end{align}
where we have dropped the corrections due to the phonon self-energy.
At the leading order in the polynomial expansion $g(p)\approx p^n b_0$:
\begin{align}
\eta\gtrsim \frac{4v^4}{25 T^2}\frac{A_{00}^2}{M_{00}}=\frac{4\mu^8}{25
v^3T^5\xi^3} \frac{\hat{A}^2_{00}}{\hat{M}_{00}} = 
\frac{4}{25 v^3}\xi^5\frac{T_F^8}{T^5} \frac{\hat{A}^2_{00}}{\hat{M}_{00}}, 
\end{align}
where we used $\mu=\xi T_F$. An interesting dimensionless
quantity to consider is the ratio of viscosity $\eta$ to the entropy
density $s$ for comparison with the conjectured bound discussed in
the introduction, Eq.~(\ref{ADS}). The phonon 
gas entropy is 
\begin{align}
s=\frac{11\pi^2}{90}\frac{T^3}{v^3},
\end{align} 
from which we obtain
\begin{align}\label{etaS}
\frac{\eta}{s}\gtrsim\frac{72}{55\pi^2}\xi^5
\frac{\hat{A}_{00}^2}{\hat{M}_{00}} 
\left(\frac{T_F}{T}\right)^8.
\end{align}

In the calculation of $\hat{M}_{st}$ in Eq.~(\ref{phase4}) the phase
space integral can be reduced to a 5-dimensional integral. Of the 
original 12-dimensional integration variables four integrations are 
removed using the energy-momentum conserving delta function $\delta^{(4)}
(p+k+p'+k')$. We choose to constrain the three-momentum $\bm{p'}$ and 
the magnitude $|\bm{k'}|$. Three more integrations can be removed as 
follows: without loss of generality we define the three-momentum 
$\bm{p} = p \hat{z}$ along the $z$-axis eliminating two angular 
integrations. Now, among the angular integration variables 
only the $z$-axis projection of the three-momenta 
$\bm{k}$ ($\hat{\bm{p}}\cdot\hat{\bm{k}} =\cos\theta$) and $\bm{k}'$
($\hat{\bm{p}}\cdot\hat{\bm{k}'} =\cos\theta'$), and the angular
separation between  $\bm{k}$ and $\bm{k}'$ ($ \hat{\bm{k}}\cdot
\hat{\bm{k}'} =\cos\gamma$) are relevant. Thus the  five remaining  
integration variables are: two magnitudes $|\bm{p}|$ and  $|\bm{k}|$, 
two angles $\theta$ and $\theta'$, and the angular difference $\phi-\phi'$.   
The 5-dimensional integration is done using the Monte Carlo routine
Vegas~\cite{Lepage:1980dq}.

%%%%%%%%%%%%%%%%%%%%%%%%%%%%%%%%%%%%%%%%%%%%%%%%%%%%%%%%%%%%%%%%%%%%
\begin{figure}[t]
\includegraphics[width=0.48\textwidth,clip]{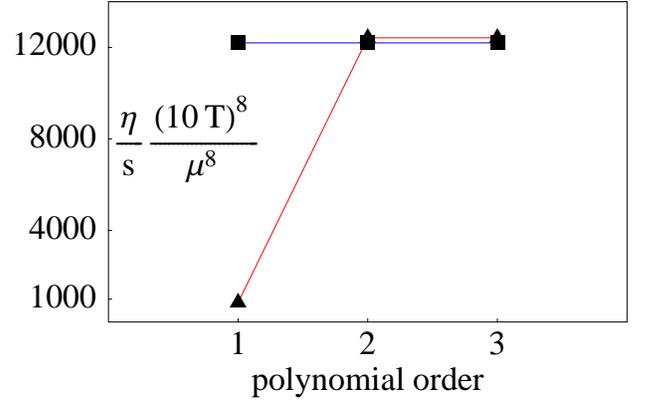}
\caption{\protect Numerical results for the scaled shear viscosity 
  to entropy density ratio $(10 T/\mu)^8\eta/s$ at $T=0.001\mu$ with
  $\xi=0.4$ as
  a function of the polynomial order $s+1$ of the polynomial $B_s(p)$. We 
  show results for two values of the variational parameter: square
  $n=-1$, triangle $n=-2$. The straight lines connecting the numerical
  results are to guide the eyes. 
}
\label{convergence}
\end{figure}  
%%%%%%%%%%%%%%%%%%%%%%%%%%%%%%%%%%%%%%%%%%%%%%%%%%%%%%%%%%%%%%%%%%%%

 In addition to varying the parameter $n$ in the trial function
$g(p)=p^n\sum_s b_s B_s(p)$, we check for convergence as we increase
the number of terms inside the summation. From numerical experiments 
with integer $n$, we find the maximal, convergent results for $n=-1$. 
In Fig.~\ref{convergence}, we show $(10 T/\mu)^8 \eta/s$ at
$T=0.001\mu$ with $\xi=0.4$  for 
$n=-2,-1$. Convergence as the order of the polynomial 
used in the trial function is varied is demonstrated. 
The $n=-2$ solution at leading 
order of the polynomial expansion starts small, and then converges to 
the $n=-1$ result. This is expected since the $n=-2$ trial function at 
second order of the polynomial $B_1(p)\sim p$ contains the trial 
function with $n=-1$. 

%%%%%%%%%%%%%%%%%%%%%%%%%%%%%%%%%%%%%%%%%%%%%%%%%%%%%%%%%%%%%%%%%%%%
\begin{figure}[t]
\includegraphics[width=0.48\textwidth,clip]{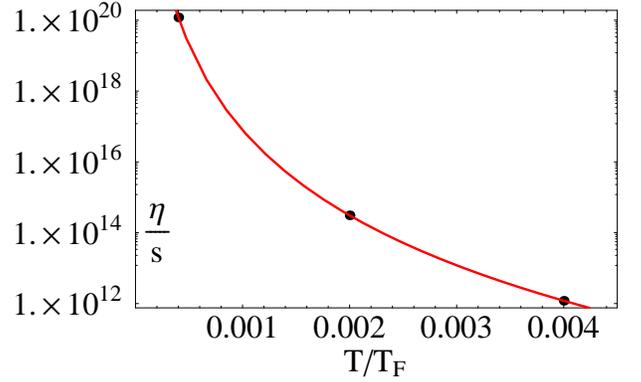}
\caption{\protect Shear viscosity to entropy density ratio as 
a function of $T/T_F$. The dots are numerical results, and the 
solid curve shows the power-law fit given in Eq.~(\ref{etaS_Fit}),
with $\xi=0.4$.   
}
\label{viscosityplotA}
\end{figure}
%%%%%%%%%%%%%%%%%%%%%%%%%%%%%%%%%%%%%%%%%%%%%%%%%%%%%%%%%%%%%%%%%%%%

\begin{figure}[t]
\includegraphics[width=0.48\textwidth,clip]{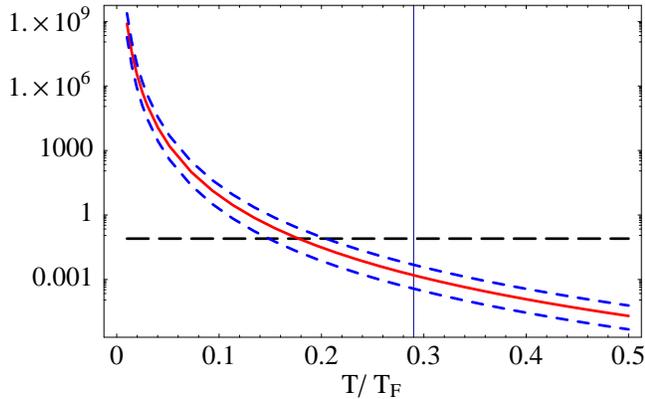}
\caption{\protect Shear viscosity to entropy density ratio $\eta/s 
  = 7.7\times 10^{-6}\xi^{5}(T_F/T)^8$ compared to the proposed
  bound $1/(4\pi)$. We show results for three values of the 
  universal parameter $\xi$: $\xi=0.4$ (solid curve), $\xi=0.3,0.5$ 
  (short-dashed curves). The critical $T_c=0.29 T_F$ is indicated. 
}
\label{viscosityplotB}
\end{figure}

Fig.~\ref{viscosityplotA} shows $\eta/s$ at three temperatures for 
the best trial function with $n=-1$. The data is very well described
by the functional form
\begin{align}\label{etaS_Fit}
\frac{\eta}{s} =&
7.7\times 10^{-6}\xi^5\frac{T_F^8}{T^8}, 
\end{align}
which is also shown in the figure. The numerical results in 
Fig.~\ref{viscosityplotA} are stable to
about $1\%$. A comparison with the conjectured
viscosity bound $1/(4\pi)$ is shown in Fig.~\ref{viscosityplotB}. The 
bound is violated for $T> 0.2 T_F$, which is close to the measured 
critical temperature $T_c=(0.29\pm0.02)T_F$~\cite{TC} for superfluidity 
where the phonon calculation is not reliable. In the region where the 
phonon calculation is reliable, the viscosity bound is satisfied.  

%%%%%%%%%%%%%%%%%%%%%%%%%%%%%%%%%%%%%%%%%%%%%%%%%%%%%%%%%%%%%%%%%%%%
%%%%%%%%%%%%%%%%%%%%%%%%%%%%%%%%%%%%%%%%%%%%%%%%%%%%%%%%%%%%%%%%%%%%

\begin{figure}[t]
\includegraphics[width=0.48\textwidth,clip]{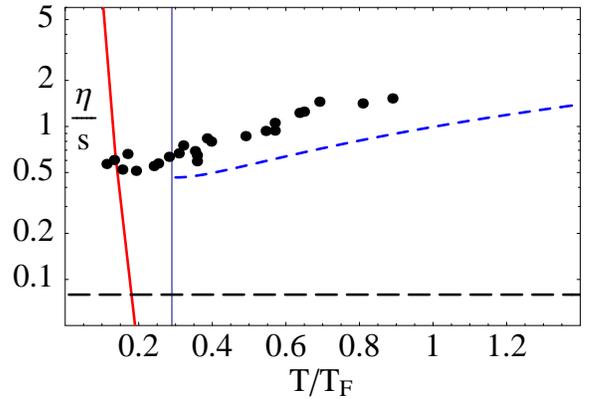}
\caption{\protect Shear viscosity to entropy density ratio $\eta/s$
  as a function of $T/T_F$. Solid curve: low temperature behavior of 
  $\eta/s$ from Eq.~(\ref{etaS_Fit}) with $\xi=0.4$, dashed curve: high temperature behavior
  of $\eta/s$ from Eqs.~(\ref{eta_aa}) and (\ref{s_highT}), long-dashed curve: proposed
  viscosity bound $1/(4\pi)$. Dots are  data from
  ~\cite{Schafer:2007pr}.  
  The critical $T_c=0.29 T_F$ is indicated.
}
\label{viscExp}
\end{figure}

 In Fig.~\ref{viscExp} we compare our results to calculations 
in the high temperature limit and to experimental data. The 
high temperature results are taken from~\cite{Bruun}. These 
authors computed the viscosity due to binary fermion collisions. 
The free space cross section is proportional to $1/k^2$. In the 
high temperature limit the infrared divergence is effectively 
cut off by the thermal momentum $(mT)^{1/2}$. For $T\gg T_c$
~\cite{Bruun}
\begin{align}
\label{eta_aa}
\eta\approx \frac{15}{32\sqrt{\pi}}(m T)^{3/2}.
\end{align} 
In this limit the entropy density is that of a classical gas
\begin{align}\label{s_highT}
s=\frac{2\sqrt{2}}{3\pi^2} (m T_F)^{3/2}\[\log\(\frac{3\sqrt{\pi}}{4} 
\frac{T^{3/2}}{T_F^{3/2}}\)+\frac{5}{2}\] .
\end{align}
The data points are based on a hydrodynamic analysis~\cite{Schafer:2007pr} 
of experimental data on the damping of collective excitations in 
a unitarity Fermi gas~\cite{Kinast:2005}. 

%%We observe that the low and high temperature curves intersect
%%at $T\simeq 0.18 T_F$, which is indeed close to the transition 
%%temperature. We also note that a crude interpolation between
%%the two limiting curves suggests that the viscosity minimum
%%is about a factor 5 above the viscosity bound. 

We observe that the na\"{i}ve extrapolation of the high $T\gg T_c$ and the low
$T\ll T_c$ curves cross at around $T\approx 0.2 T_F$, which is indeed
close to the transition temperature $T_c\approx 0.29 T_F$. This crude
extrapolation of the two limiting curves for $\eta/s$ suggests that 
the viscosity minimum
is about a factor 5 above the viscosity bound. 
This is quite 
consistent with the experimental data. We also note that the 
experimental data show the expected increase in $\eta/s$ for
$T\gg T_c$, but they do not show the rise for $T\ll T_c$. This
may be related to the fact that the phonon mean free path 
becomes so large that it is comparable to the size of the experimental
Fermi gas sample and hydrodynamics does not apply.

%%%%%%%%%%%%%%%%%%%%%%%%%%%%%%%%%%%%%%%%%%%%%%%%%%%%%%%%%%%%%%%%%%%%
%%%%%%%%%%%%%%%%%%%%%%%%%%%%%%%%%%%%%%%%%%%%%%%%%%%%%%%%%%%%%%%%%%%%

%========================================================================
\section{Conclusions}
\label{conclude}
%========================================================================

 We computed the shear viscosity of a cold unitarity gas in the 
superfluid phase. For $T\ll T_c\sim T_F$ the viscosity is dominated 
by phonons, and the leading order effective Lagrangian for 
the phonons is characterized by a single universal parameter 
$\xi$. This parameter can be extracted from the ground state
energy of the  unitarity gas. 

 The calculation is based on the linearized Boltzmann equation,
and only the leading order $2\leftrightarrow 2$ phonon scattering processes
are included. The shear viscosity is determined using a variational 
procedure. We find that the shear viscosity scales as 
$\eta \gtrsim 9.3 \times 10^{-6}
\xi^5 T_F^8/(v^3 T^5)$. This result can be combined with high
temperature calculations of the shear viscosity to provide an
estimate of the location and magnitude of the viscosity minimum.
We find that the minimum value of $\eta/s$ occurs close to $T_c$,
and that the value of $\eta/s$ is likely to exceed the proposed
viscosity bound. A similar viscosity minimum is expected to occur
in QCD. At low temperature the viscosity is dominated
by weakly interacting Nambu-Goldstone bosons (pions and kaons)
\cite{Prakash:1993bt,Dobado:2001jf,Chen:2006ig}, and at high
temperature the viscosity is governed by weakly
interacting quarks and gluons \cite{Arnold:2003zc}.

 There are a number of issues that deserve further study. The
viscosity of the superfluid unitarity gas has the same $1/T^5$ 
behavior as the viscosity of liquid Helium at low temperature. 
In the case of liquid Helium the viscosity is believed to be 
dominated by $1\leftrightarrow 2$ processes. On-shell phonon
splitting processes can only occur if higher order corrections
to the effective Lagrangian lead to a concave phonon dispersion 
relation [$E_p=v|\bm{p}|(1+\gamma {\bm{p}}^2$) with $\gamma>0$]. 
For the unitarity Fermi gas we do not know whether this is the case. 
It is known that the Bogoliubov spectrum of a weakly interacting 
Bose gas has $\gamma>0$, so the role of these processes can 
be studied in an expansion around the Bose-Einstein limit.

 In general we would like to extend the calculation to higher
temperatures. In the vicinity of the $T_c$ we expect both 
bosonic and fermionic excitations to play a role. A possible
starting point in this regime is provided by the expansion
around $d=4-\epsilon$ spatial dimensions proposed in 
\cite{Nussinov:2004,Nishida:2006br}. It is also interesting
to improve the high temperature calculations by including
correlations between the fermions. Some steps in this direction
were taken in \cite{Bruun}.

\begin{acknowledgments}
The authors acknowledge helpful discussions with Dean Lee. 
This work was supported in parts by the US Department of Energy grant 
DE-FG02-03ER41260. 
\end{acknowledgments}

%========================================================================
\appendix
\section{Analytic form for the self-energy}
\label{sigmaAnalytic}
%========================================================================

 The imaginary part of $\Sigma(p_0, \bm{p})$ arises from the pole
terms in Eq.~(\ref{sigma}). We find (see \cite{Manuel:2004iv})
\begin{align}
&\operatorname{Im}\Sigma(p_0,\bm{p})
= \frac{\pi^3\xi^{3/2}}{16\sqrt{3}\mu^4} \sum_{s_1,s_2=\pm}
\int\frac{d^3 k}{(2\pi)^3}
\frac{s_1 s_2}{4 E_k E_{\bm{p-k}}}H(P,K,s_1)\nonumber\\
& \hspace{0.5cm}\times\[1+f^{(0)}_{s_1 E_k}
 +f^{(0)}_{s_2 E_{\bm{p-k}}}\]\delta(p_0-s_1 E_k-s_2
 E_{\bm{p-k}}),
\end{align}
with
\begin{align}
H(P,K;s)\equiv&\[p_0 (2P\cdot K -K^2)+k_0 (P^2-2P\cdot K)\]^2\Big|_{k_0=s
 E_k}.
\end{align}
There are four terms, corresponding to $s_1,s_2=\pm 1$. Terms with
$s_1\neq s_2$ contribute for space-like momenta $v|\bm{p}|>p_0$,
and terms with $s_1= s_2$ contribute for time-like momenta. For
space-like momenta we get
\begin{align}
&\operatorname{Im}\Sigma(p_0,\bm{p})
 = \frac{3 \sqrt{3}\pi\xi^{3/2}p_0^2}{128 v |\bm{p}|\mu^4}
   \int_{\frac{p_0-v|\bm{p}|}{2}}^{\infty} d|\bm{k}|\, \\
 & \hspace{0.25cm}
    \left(8v^2 {\bm{k}}^2-8p_0 v|\bm{k}|-3v^2 {\bm{p}}^2+3 p_0^2\right)^2
    \left(f^{(0)}_{|v\bm{k}|}-f^{(0)}_{v|\bm{k}|-p_0}\right). \nonumber
\end{align}
The result for time-like momenta and $p_0>0$ is
\begin{align}
&\operatorname{Im}\Sigma(p_0,\bm{p})
 = \frac{3 \sqrt{3}\pi\xi^{3/2}p_0^2}{256v |\bm{p}|\mu^4}
   \int_{\frac{p_0-v|\bm{p}|}{2}}^{\frac{p_0+v|\bm{p}|}{2}} d|\bm{k}|\, \\
 & \hspace{0.25cm}
    \left(8v^2 {\bm{k}}^2-8p_0 v|\bm{k}|-3v^2 {\bm{p}}^2+3 p_0^2\right)^2
    \left(1+f^{(0)}_{v|\bm{k}|}-f^{(0)}_{p_0-v|\bm{k}|}\right), \nonumber
\end{align}
and $\operatorname{Im}\Sigma(-p_0,\bm{p})=-\operatorname{Im}\Sigma(p_0,
\bm{p})$. These integrals can be computed analytically in the limit
of small momenta. In the space-like region we have $|p_0|\leq v |\bm{p}|
\ll v |\bm{k}|\sim T$. This implies $f^{(0)}_{v|\bm{k}|}-
f^{(0)}_{v|\bm{k}|-p_0} \approx p_0 (f^{(0)}_{v|\bm{k}|})'$ and
\begin{align}
\operatorname{Im}\Sigma(p_0,\bm{p})\approx
\frac{2\sqrt{3}\pi^5}{5\mu^4 v}\xi^{3/2}T^4\frac{p_0^3}{|\bm{p}|}
\Theta(v^2 |\bm{p}|^2-p_0^2) .
\end{align}
For time-like momenta $|p_0|\sim v |\bm{k}|\sim T\gg v |\bm{p}|$, and
\begin{align}
\operatorname{Im}\Sigma(p_0,\bm{p})&\approx
\frac{3\sqrt{3}\pi}{256}\xi^{3/2}
p_0^6
\[\frac{\exp(\frac{p_0}{2 T})+1}{\exp(\frac{p_0}{2 T})-1}\Theta(p_0)\right.\\
&\left.-
\frac{\exp(\frac{-p_0}{2 T})+1}{\exp(\frac{-p_0}{2 T})-1}\Theta(-p_0)
\].\nonumber
\end{align}

%==============Bibliography =====================

%\bibliographystyle{h-physrev4}
%\bibliography{PhononsUnitarity.bib}

\begin{thebibliography}{10}

\bibitem{Danielewicz:1984ww}
P.~Danielewicz and M.~Gyulassy,
\newblock Phys. Rev. {\bf D31}, 53 (1985).
%%CITATION = PHRVA,D31,53;%%

\bibitem{Policastro:2001yc}
G.~Policastro, D.~T. Son and A.~O. Starinets,
\newblock Phys. Rev. Lett. {\bf 87}, 081601 (2001), [hep-th/0104066].
%%CITATION = HEP-TH/0104066;%%

\bibitem{Kovtun:2004de}
P.~Kovtun, D.~T. Son and A.~O. Starinets,
\newblock Phys. Rev. Lett. {\bf 94}, 111601 (2005), [hep-th/0405231].
%%CITATION = HEP-TH/0405231;%%

\bibitem{Teaney:2003kp}
D.~Teaney,
\newblock Phys. Rev. {\bf C68}, 034913 (2003), [nucl-th/0301099].
%%CITATION = NUCL-TH/0301099;%%

\bibitem{Hirano:2005wx}
T.~Hirano and M.~Gyulassy,
\newblock Nucl. Phys. {\bf A769}, 71 (2006), [nucl-th/0506049].
%%CITATION = NUCL-TH/0506049;%%

\bibitem{Cohen:2007qr}
T.~D. Cohen,
\newblock hep-th/0702136.
%%CITATION = HEP-TH/0702136;%%

\bibitem{Dobado:2007tm}
A.~Dobado and F.~J. Llanes-Estrada,
\newblock hep-th/0703132.
%%CITATION = HEP-TH/0703132;%%

\bibitem{JEThomas}
K. M. O'Hara, S. L. Hemmer, M. E. Gehm, S. R. Granade and J. E. Thomas, {\it
  Science} {\bf 298}, 2179 (2002); M. E. Gehm, S. L. Hemmer, S. R. Granade, K.
  M. O'Hara and J. E. Thomas, \Journal{\PRA}{68}{011401(R)}{2003}.

\bibitem{TBourdel}
T. Bourdel, J. Cubizolles, L. Khaykovich, K. M. F. Magalh\~{a}es, S. J. J. M.
  F. Kokkelmans, G. V. Shlyapnikov and C. Salomon,
  \Journal{\PRL}{91}{020402}{2003}.

\bibitem{KDieckmann}
K. Dieckmann, C. A. Stan, S. Gupta, Z. Hadzibabic, C. H. Schunck and W.
  Ketterle, \Journal{\PRL}{89}{203201}{2002}.

\bibitem{DJin}
C. A. Regal, C. Ticknor, J. L. Bohn and D. S. Jin, {\it Nature} {\bf 424}, 47
  (2003); M. Greiner, C. A. Regal and D. S. Jin, {\it Nature} {\bf 426}, 537
  (2003).

\bibitem{SJochim}
S. Jochim, M. Bartenstein, A. Altmeyer, G. Hendl, S. Riedl, C. Chin, J. Hecker
  Denschlag, and R. Grimm, {\it Science} {\bf 302}, 2101 (2003).

\bibitem{Bruun}
P. Massignan, G. M. Brunn and H. Smith, \Journal{\PRA}{71}{033607}{2005}; G. M.
  Bruun and H. Smith, \Journal{\PRA}{72}{043605}{2005},
  \Journal{\PRA}{75}{043612}{2007}.

\bibitem{Khalatnikov}
I.~M. Khalatnikov,
\newblock {\em An Introduction to the Theory of Superfluidity} (W.~A.~Benjamin,
  Inc., 1965).

\bibitem{Maris}
H.~J. Maris,
\newblock Phys Rev {\bf A}, 1980 (1973).

\bibitem{Manuel:2004iv}
C.~Manuel, A.~Dobado and F.~J. Llanes-Estrada,
\newblock JHEP {\bf 09}, 076 (2005), [hep-ph/0406058].
%%CITATION = HEP-PH 0406058;%%

\bibitem{Landau}
E.~M. Lifshitz and L.~P. Pitaevskii,
\newblock {\em Physical Kinetics} (Elsevier, 1987).

\bibitem{Greiter:1989qb}
M.~Greiter, F.~Wilczek and E.~Witten,
\newblock Mod. Phys. Lett. {\bf B3}, 903 (1989).
%%CITATION = MPLAE,B3,903;%%

\bibitem{Son:2005rv}
D.~T. Son and M.~Wingate,
\newblock Annals Phys. {\bf 321}, 197 (2006), [cond-mat/0509786].
%%CITATION = COND-MAT/0509786;%%

\bibitem{Son:2002zn}
D.~T. Son,
\newblock hep-ph/0204199.
%%CITATION = HEP-PH 0204199;%%

\bibitem{Carlson2003}
J.~Carlson, S.-Y. Chang, V.~R. Pandharipande and K.~E. Schmidt,
\newblock Phys. Rev. Lett. {\bf 91}, 050401 (2003).

\bibitem{Chang2004}
S.~Y. Chang, V.~R. Pandharipande, J.~Carlson and K.~E. Schmidt,
\newblock Phys. Rev. A {\bf 70}, 043602 (2004).

\bibitem{Astrakharchik2004}
G.~E. Astrakharchik, J.~Boronat, J.~Casulleras and S.~Giorgini,
\newblock Phys. Rev. Lett. {\bf 93}, 200404 (2004).

\bibitem{Chen:2003vy}
J.~W. Chen and D.~B. Kaplan,
\newblock Phys. Rev. Lett. {\bf 92}, 257002 (2004).
%%CITATION = HEP-LAT 0308016;%%

\bibitem{Lee:2005fk}
D.~Lee,
\newblock Phys. Rev. {\bf B73}, 115112 (2006).
%%CITATION = COND-MAT 0511332;%%

\bibitem{Bulgac:2005pj}
A.~Bulgac, J.~E. Drut and P.~Magierski,
\newblock cond-mat/0505374.
%%CITATION = COND-MAT 0505374;%%

\bibitem{Burovski:2006}
E.~Burovski, N.~Prokof'ev, B.~Svistunov and M.~Troyer,
\newblock Phys. Rev. Lett. {\bf 96}, 160402 (2006).

\bibitem{TC}
L.~Luo, B.~Clancy, J.~Joseph, J.~Kinast and J.~E. Thomas,
\newblock Phys. Rev. Lett. {\bf 98}, 080402 (2007), [cond-mat/0611566].
%%CITATION = COND-MAT 0611566;%%

\bibitem{Resibois}
P.~R\'{e}sibois and M.~d. Leener,
\newblock {\em Classical Kinetic Theory of Fluids} (John Wiley $\&$ Sons,
  1977).

\bibitem{Chen:2006ig}
J.-W. Chen and E.~Nakano,
\newblock Phys. Lett. {\bf B647}, 371 (2007), [hep-ph/0604138].
%%CITATION = HEP-PH/0604138;%%

\bibitem{Dobado:2001jf}
A.~Dobado and S.~N. Santalla,
\newblock Phys. Rev. {\bf D65}, 096011 (2002), [hep-ph/0112299].
%%CITATION = HEP-PH 0112299;%%

\bibitem{Lepage:1980dq}
G.~P. Lepage,
\newblock CLNS-80/447.

\bibitem{Schafer:2007pr}
T.~Schafer,
\newblock cond-mat/0701251.
%%CITATION = COND-MAT/0701251;%%

\bibitem{Kinast:2005}
J.~Kinast, T.~A. and T.~J. E.,
\newblock Phys. Rev. Lett. {\bf 94}, 170404 (2005), [cond-mat/0502507].
%%CITATION = COND-MAT/0502507;%%

\bibitem{Prakash:1993bt}
M.~Prakash, M.~Prakash, R.~Venugopalan and G.~Welke,
\newblock Phys. Rept. {\bf 227}, 321 (1993).
%%CITATION = PRPLC,227,321;%%

\bibitem{Arnold:2003zc}
P.~Arnold, G.~D. Moore and L.~G. Yaffe,
\newblock JHEP {\bf 05}, 051 (2003), [hep-ph/0302165].
%%CITATION = HEP-PH/0302165;%%

\bibitem{Nussinov:2004}
Z.~Nussinov and S.~Nussinov,
\newblock cond-mat/0410597.
%%CITATION = COND-MAT/0410597;%%

\bibitem{Nishida:2006br}
Y.~Nishida and D.~T. Son,
\newblock Phys. Rev. Lett. {\bf 97}, 050403 (2006), [cond-mat/0604500].
%%CITATION = COND-MAT/0604500;%%

\end{thebibliography}

%==================================  End ================================
\end{document}